\documentclass[useAMS,usenatbib]{mn2e}  
\usepackage{graphicx}
\usepackage{txfonts}
\usepackage{natbib}
\usepackage{color}
\bibpunct{(}{)}{;}{a}{}{,}
\title{3D simulations of disc-winds extending radially self-similar MHD models}
\author[M. Stute et al.]{Matthias 
Stute,$^{1}$\thanks{E-mail:matthias.stute@uni-tuebingen.de} 
Jos\'e Gracia,$^{2}$ Nektarios Vlahakis,$^{3}$ Kanaris Tsinganos,$^{3,4}$ 
\newauthor Andrea Mignone$^{5}$ and Silvano Massaglia$^{5}$ \\
$^{1}$Institute of Astronomy and Astrophysics, Section Computational Physics, 
Eberhard Karls Universit\"at T\"ubingen, Auf der Morgenstelle 10, 72076 
T\"ubingen, Germany \\
$^{2}$High Performance Computing Center Stuttgart (HLRS), Universit\"at 
Stuttgart, 70550 Stuttgart, Germany \\ 
$^{3}$Department of Astrophysics, Astronomy and Mechanics,
Faculty of Physics, University of Athens, 15784 Zografos, Athens, Greece
\\ 
$^{4}$
National Observatory of Athens, Lofos Nymphon, Thission 11810, Athens, Greece 
\\
$^{5}$Dipartimento di Fisica, Universit\`a degli Studi di Torino, via Pietro 
Giuria 1, 10125 Torino, Italy}

\begin{document}

\date{Accepted 2014 January 28. Received 2014 January 27; in original form 2013 November 04}

\pagerange{\pageref{firstpage}--\pageref{lastpage}} \pubyear{2014}

\maketitle

\label{firstpage}

\begin{abstract}
Disc-winds originating from the inner parts of accretion 
discs are considered as the basic component of magnetically collimated 
outflows. The only available analytical MHD solutions to describe 
disc-driven jets are those characterized by the symmetry of radial 
self-similarity. 
However, radially self-similar MHD jet models, in general, have three 
geometrical shortcomings, (i) a singularity at the jet axis, (ii) the necessary 
assumption of axisymmetry, and (iii) the non-existence of an intrinsic 
radial scale, i.e. the jets formally extend to radial infinity. 
Hence, numerical simulations are necessary to extend the analytical 
solutions towards the axis, by solving the full three-dimensional equations of 
MHD and impose a termination radius at finite radial distance.
We focus here on studying the effects of relaxing the (ii) assumption of 
axisymmetry, i.e. of performing full 3D numerical simulations of a 
disc-wind crossing all magnetohydrodynamic critical surfaces. We compare 
the results of these runs with previous axisymmetric 2.5D simulations.
The structure of the flow in all simulations shows strong similarities. The
3D runs reach a steady state and stay close to axisymmetry for most of 
the physical quantities, except for the poloidal magnetic field and the 
toroidal velocity which slightly deviate from axisymmetry. 
The latter quantities show signs of instabilities, which, however, are confined 
to the region inside the fast magnetosonic separatrix surface. The forces 
present in the flow, both of collimating and accelerating nature, are in 
good agreement in both the 2.5D and the 3D runs.
We conclude that the analytical solution behaves well also after relaxing the 
basic assumption of axisymmetry. 
\end{abstract}
    

\section{Introduction}

Astrophysical jets are observed in association with a wide range of objects, 
from Brown Dwarfs and young stellar objects to supermassive Black Holes in 
active galactic nuclei; however, there are still open several questions 
concerning the launching and acceleration mechanisms of jets. In all cases, 
jets and discs are inter-related, while magnetic fields play a key role in 
accelerating the outflows. \citet{BlP82} studied the magneto-centrifugal 
acceleration along magnetic field lines threading an accretion disc. They showed
the braking of matter in the azimuthal direction inside the disc and the outflow
acceleration above the disc surface guided by the poloidal magnetic field 
components. Toroidal components of the magnetic field then collimate the 
outflow. Numerous semi-analytic models extended the work of Blandford \& Payne 
along the guidelines of radially self-similar solutions of the full 
magnetohydrodynamic (MHD) equations 
\cite[e.g.][]{CoL94, Li95, Li96, Fer97, VlT98}. 
Several numerical studies exist, which have focused on the magnetic launching of
disc-winds. In most models a polytropic equilibrium accretion disc has been 
used as a boundary condition 
\citep[e.g.][]{UKR95, KLB99, KLB03, OCP03, NaM04, ALK05, ALK06, PRO06}. 
The magnetic feedback on the disc structure was therefore not calculated 
self-consistently. Only in recent years have been carried out the first 
simulations which include self-consistently the accretion disc in the 
calculations of jet launching 
\citep[e.g.][]{CaK02, CaK04, KMS04, ZFR07, TFM09, MFZ10, SFP12, TFM13, FeS13}. 

Often analytical solutions and their integrals of motion are used for connecting
observed quantities far away from the jet source with properties of the 
jet-driving accretion disc. The behaviour of integrals of motions has been also 
tested in numerical simulations. Several 2.5D simulations of disc winds from 
the disc as a boundary condition \citep[e.g.][]{RUK97, OuP97, KLB99, UKR99} 
found super-fast magnetosonic flows with properties which are expected from 
self-similar theory, although they have not used self-consistent analytical 
solutions for describing the boundary conditions. Recently, \citet{SNO10} 
performed 3D simulations extending the calculations of \citet{OuP97} and found 
good agreement between both. In addition, \citet{SNO10} also simulated 
configurations closer to self-similar solutions (their run BP) and compared 
synthetic emission maps derived from them with HST observations. A similar 
comparison is presented in \citet[][hereafter paper~II]{SGT10} using 
simulations from \citet[][hereafter paper~I]{STV08}.

Most self-similar models have three serious
limitations, (1) the flow often does not cross all critical points (especially 
not the fast-magnetosonic limiting characteristic), with the result that the 
terminal wind solution is not causally disconnected from the disc, (2) 
singularities exist at the jet axis in radial self-similar models, and (3) for 
deriving self-similar models the assumption of axisymmetry is necessary. 

\citet[][V00 hereafter]{VTS00} showed that a terminal wind solution can be 
constructed that is causally disconnected from the disc and hence any 
perturbation downstream of the superfast transition cannot affect the upstream 
structure of the steady outflow. The other two limitations can only be solved 
using numerical simulations extending the analytical solution of e.g. V00 as 
done by \citet[]{GVT06} using the MHD code NIRVANA 
\citep{Zie98}, \citet[]{MTV08} using the MHD code
PLUTO\footnote{publicly available at http://plutocode.ph.astro.it/} 
\citep{MBM07} and again with PLUTO in paper I for comparison with 
models with finite jet-emitting disc radii. 
\citet{CGV08} extended the solution by adding the effects of resistivity. 
However, all of those models still assumed axisymmetry.

The aim of this paper is to investigate numerically how relaxing the assumption 
of axisymmetry, i.e. performing full 3D numerical simulations of a disc-wind 
crossing all magnetohydrodynamic critical surfaces, affects the topology, 
structure and stability of this particular radial self-similar analytical 
solution.

This paper is organized as follows: the numerical setup is briefly described in 
Sec.~\ref{sec_num_models}. In Sec.~\ref{sec_results} we describe the results of 
our simulation. We close with a summary and discussion of the results in the 
last section.

\section{The numerical setup} \label{sec_num_models}

The time-dependent, ideal MHD equations to be solved 
numerically are:
\begin{eqnarray}
\frac{\partial \rho}{\partial t} + \nabla\cdot(\rho\,\bmath{v}) &=& 0 \,, \\
\frac{\partial \bmath{v}}{\partial t} + (\bmath{v} \cdot \nabla)\,\bmath{v} - 
\frac{1}{\rho}\,(\nabla\times\bmath{B})\times \bmath{B} + \frac{1}{\rho}\,
\nabla\,p &=& -\nabla\Phi \,, \\
\label{energy}
\frac{\partial p}{\partial t} + \bmath{v}\cdot\nabla\,p + 
\Gamma\,p\,\nabla\cdot\bmath{v} &=&
\Lambda \,, \\
\frac{\partial \bmath{B}}{\partial t} - \nabla\times(\bmath{v}\times\bmath{B}) 
&=& 0
\,, \\ \nabla\cdot \bmath{B} &=& 0 \,,
\end{eqnarray}
where $\rho$, $p$, $\bmath{v}$, $\bmath{B}$ denote the density, pressure, 
velocity and magnetic field over $\sqrt{4\,\pi}$, respectively. 
$\Phi = -\mathcal{G}\,\mathcal{M} / r$ is the gravitational potential of the 
central object with mass $\mathcal{M}$, $\Lambda$ represents the rate of the 
volumetric energy gain/loss terms ($\Lambda = [\Gamma - 1]\,\rho\,q$, with $q$ 
the rate of the energy gain/loss terms per unit mass), and $\Gamma$ is the 
ratio of the specific heats. The spherical radius is denoted by $r$ and the 
cylindrical radius by $R$.

As initial conditions, we use the steady, radially self-similar solution which 
is described in V00 and crosses all three appropriate critical surfaces,
modified near the symmetry axis as described in \cite{GVT06} and \cite{MTV08}. 
We note that a polytropic relation between the density and the pressure is 
assumed, i.e. $P = Q(A)\,\rho^\gamma$, with $\gamma$ being the effective 
polytropic index and A the magnetic flux function. Equivalently, the source 
term in Eq.~(\ref{energy}) has the special form 
 \begin{equation}
\Lambda = (\Gamma - \gamma)\,p\,(\nabla\cdot\bmath{v}) \,,
\end{equation}
transforming the energy Eq.~(\ref{energy}) to
\begin{equation}
\frac{\partial p}{\partial t} + \bmath{v}\cdot\nabla\,p + 
\gamma\,p\,\nabla\cdot\bmath{v} = 0 \,.
\end{equation}

The latter can be interpreted as describing the adiabatic evolution of a gas 
with ratio of specific heats $\gamma$, whose ``effective entropy'' 
$P / \rho^\gamma$ is conserved along each streamline. We refer the reader 
to V00 and paper~I for further details on the self-similar solution and the 
numerical setup. We define the reference length $R_*$ to be unity, while the 
reference velocity is normalized by setting $v_* = 1$. Time is given in units 
of $t_0 = 2\,\pi\,\sqrt{R_*^3 / \mathcal{G}\,\mathcal{M}}$, i.e. one Keplerian 
orbit period at $R_* = 1$.

\begin{figure*}
  \centering
  \includegraphics[width=\textwidth]{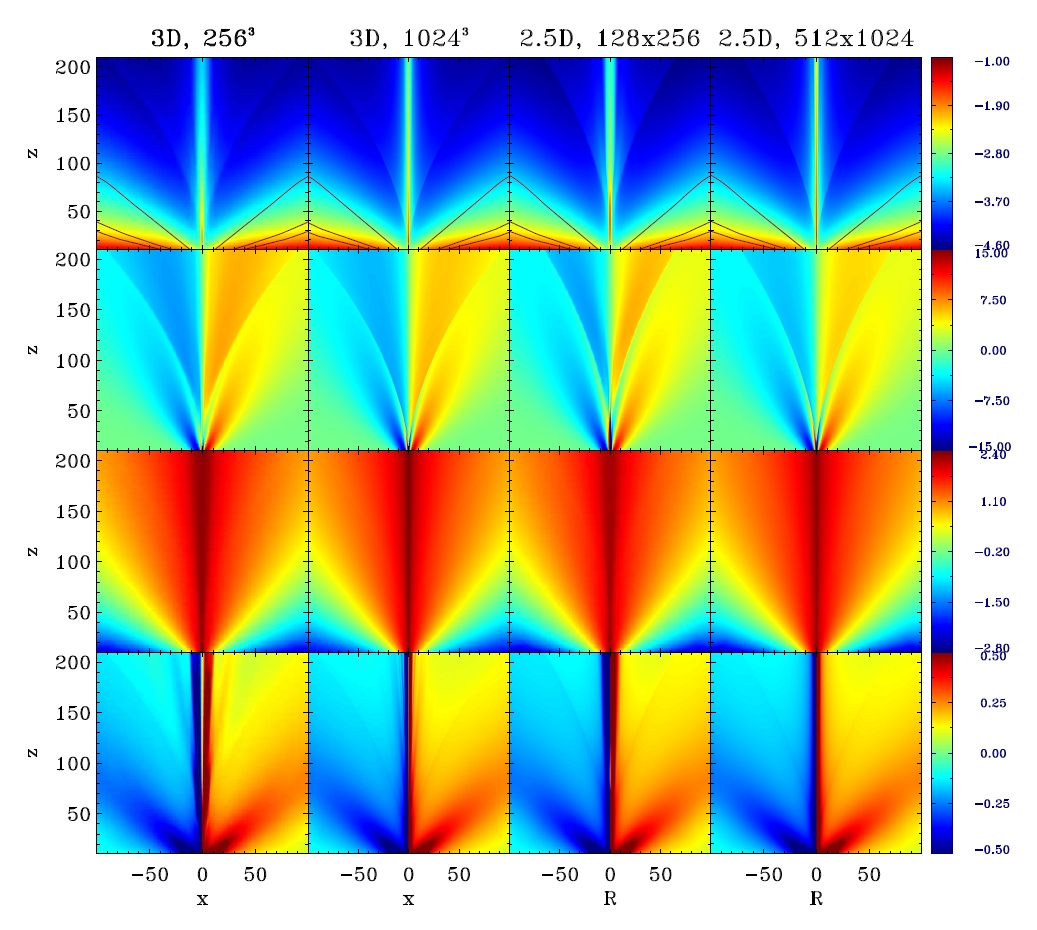}
  \caption{Maps of $\log ( \rho )$, $v_R$, $\log ( v_z )$ and $v_{\phi}$ of the 
    final state (from top to bottom) for the two 3D runs in the 
    $x$-$z$-plane and two axisymmetric 2.5D runs which have been mirrored 
    along the $z$ axis for comparison. In the density maps, we also plot the 
    three surfaces (black lines) where the flow velocity exceeds the 
    slow-magnetosonic, the Alfv\'en and the fast-magnetosonic speed, 
    respectively. Note the discontinuity of the physical quantities along the 
    FMSS which protects the sub-FMSS region from the perturbation arising from 
    interpolating the solution to avoid the singularity at the axis.  In other 
    words, the only real critical surface is the FMSS through which any 
    perturbation arising downstream cannot propagate upstream toward the base 
    of the solution. No discontinuity of any physical quantity is seen along 
    the slow, Alfv\'en and fast surfaces.}
  \label{Fig_struct_models}
\end{figure*}

We solve the MHD equations with the PLUTO code, a modular Godunov-type code 
particularly oriented towards the treatment of astrophysical flows in the 
presence of discontinuities. For the present case, second order accuracy is 
achieved using a Runge-Kutta scheme (for temporal integration) and piecewise 
linear reconstruction (in space). All the computations were carried out with 
the Harten, Lax, Van Leer approximate Riemann Solver with the contact 
discontinuity (HLLC).

At the lower boundary, we keep the physical variables fixed to their analytical 
values, however, making sure that the problem is not over-specified. Outflow 
conditions are set at the other boundaries, i.e. all gradients across these 
boundaries are set to zero. 

We performed 3D simulations on grids with resolutions between 256$^3$ and 
1024$^3$ cells and a domain size of [-100,100]x[-100,100]x[10,210] in Cartesian 
coordinates ($x$,$y$,$z$). For comparison, we also run 2.5D axisymmetric 
simulations with the setup taken from paper~I, but now with a domain size of 
[0,100]x[10,210] in cylindrical coordinates ($R$,$z$) and grids with resolutions
between 128x256 cells and 512x1024 cells, i.e. with the same resolutions as the 
3D runs.

\section{Results of the simulations} \label{sec_results}

\subsection{Basic structure of the flow}

The three-dimensional and 2.5-dimensional simulations show a similar structure 
of the flow. Since the initial conditions are not a steady solution of the 
system of equations under consideration due to the modifications at the axis, 
the initial conditions relax toward a new final steady state within about 
$10\,t_0$, while after that only changes of a few percent are present. As 
expected from the high MHD signal velocities, the inner regions of the flow 
evolve very rapidly. MHD waves communicate changes of the inner flow to the 
outer regions and are clearly visible as bends moving along the field lines. In 
all models, a shock representing the fast magnetosonic separatrix surface 
(FMSS) forms, as already described in the axisymmetric 2.5D models by 
\cite{GVT06}, \cite{MTV08} and paper~I.

In Fig. \ref{Fig_struct_models}, we plot logarithmic maps of the final state 
for the density and velocity components and also the position of the three 
surfaces where the flow velocity exceeds the slow-magnetosonic, the 
Alfv\'en and the fast-magnetosonic speed, respectively. All eight physical 
quantities (including pressure and magnetic field components) agree well in the 
axisymmetric 2.5D and the 3D models with the same resolution. Since the central 
axis is not a symmetry axis anymore in the 3D simulations, some differences in 
the region $R < 10$ are present. 

\subsection{Deviations from axisymmetry}

\begin{figure*}
  \centering
  \includegraphics[width=0.65\columnwidth]{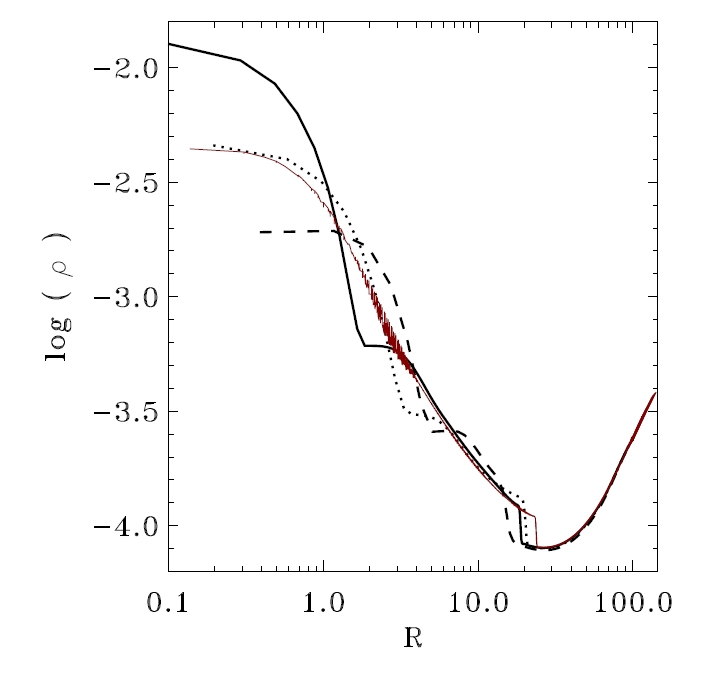}
  \includegraphics[width=0.65\columnwidth]{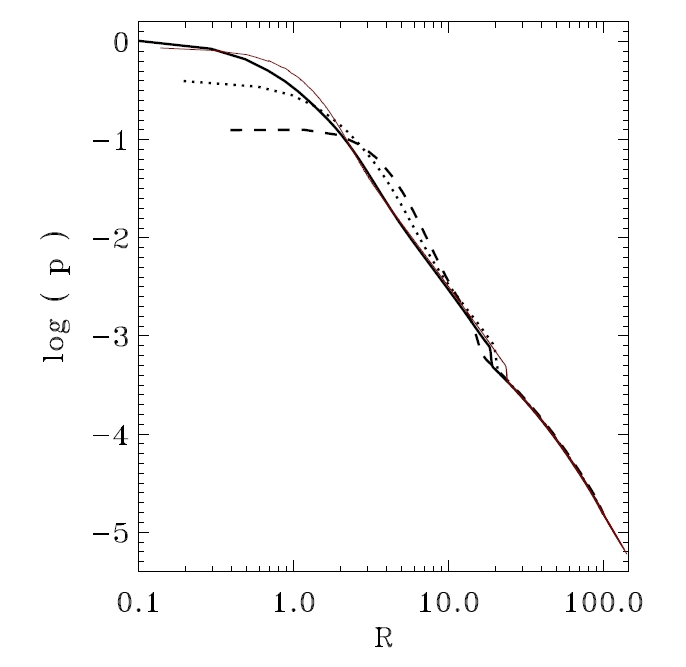} \\
  \includegraphics[width=0.65\columnwidth]{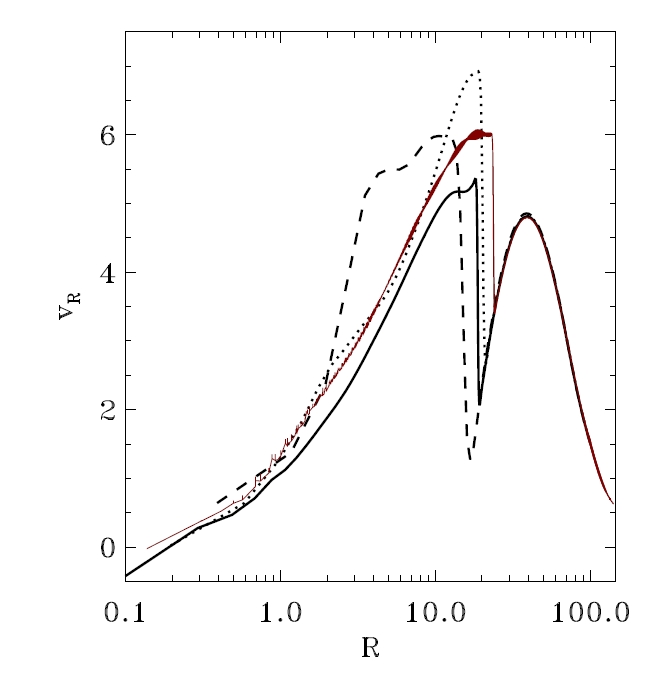}
  \includegraphics[width=0.65\columnwidth]{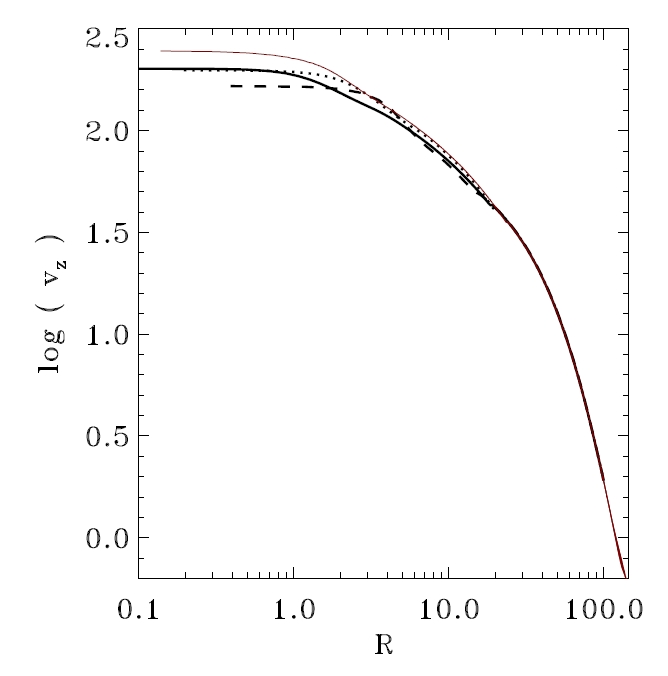}
  \includegraphics[width=0.65\columnwidth]{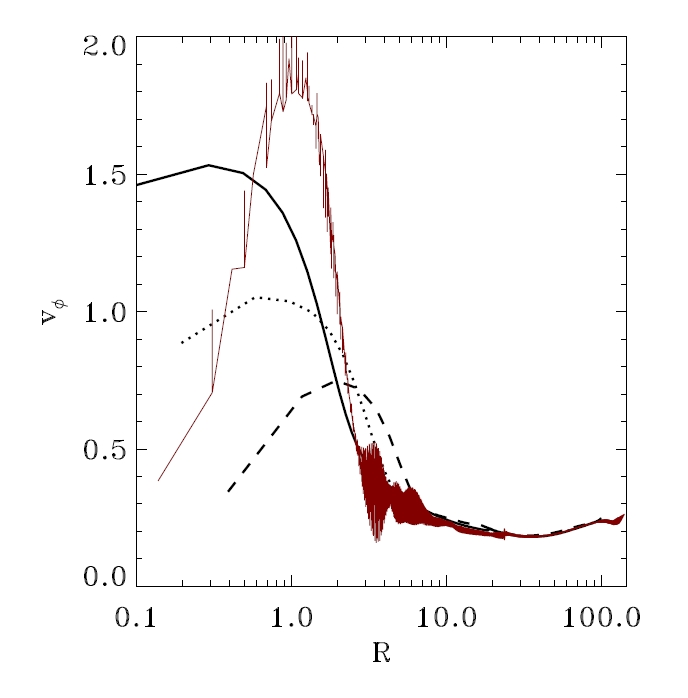} \\
  \includegraphics[width=0.65\columnwidth]{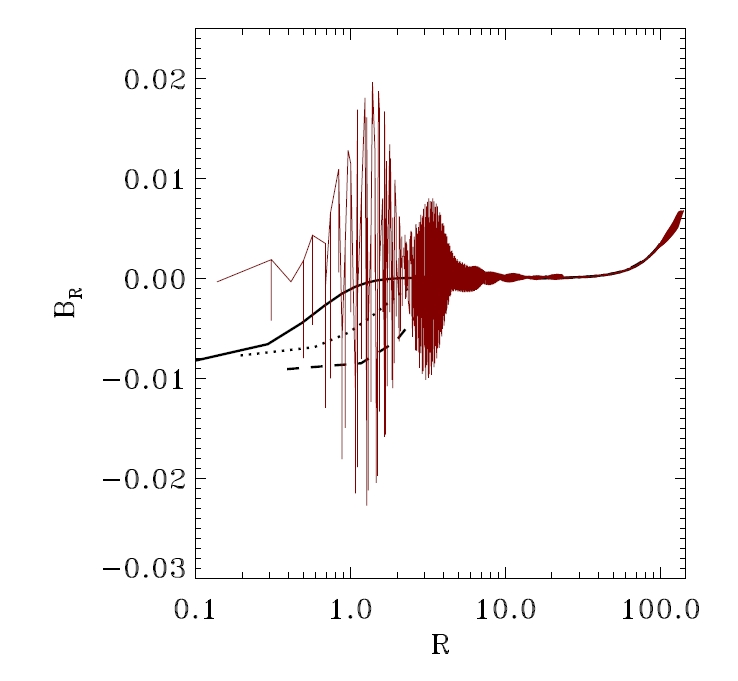}
  \includegraphics[width=0.65\columnwidth]{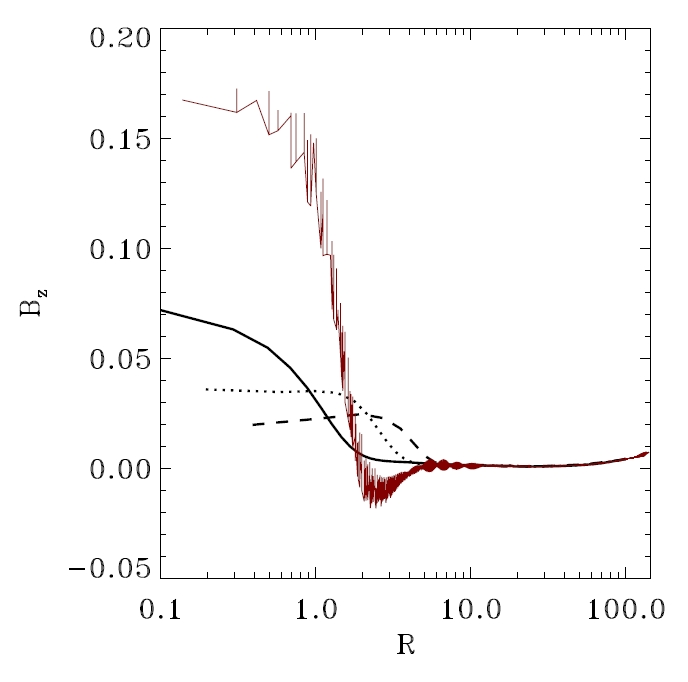}
  \includegraphics[width=0.65\columnwidth]{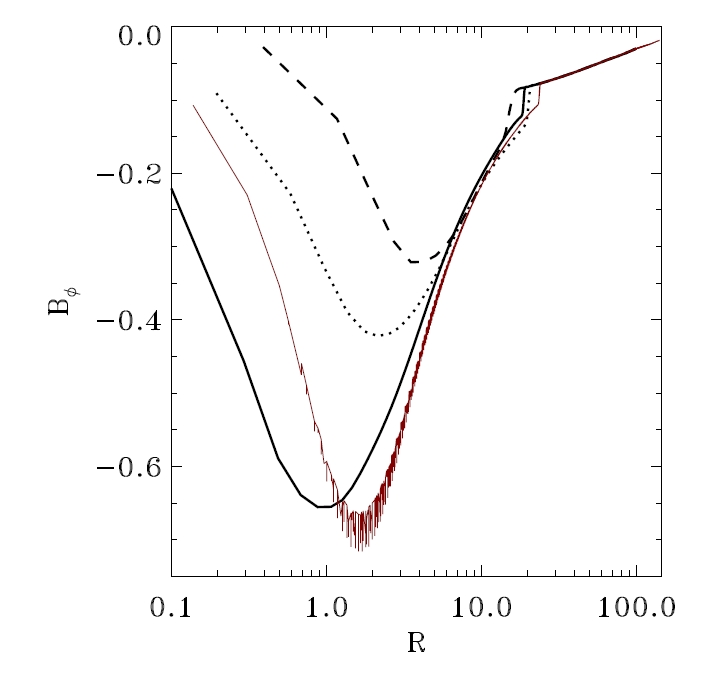}
  \caption{Profiles of all quantities as a function of cylindrical radius at 
    $z = 100$ in the 2.5D runs with 128$\times$256 grid cells (dashed), 
    256$\times$512 grid cells (dotted) and 512$\times$1024 grid cells (solid) 
    and in the finer 3D run (red points). Note that in the 3D run (a) $v_R$, 
    $v_\phi$, $B_R$, and $B_\phi$ go through zero near $R=0$, meaning that the 
    rotation axis remains stable during the simulation. (b) $\rho$, $p$, 
    $v_R$, $v_\phi$ and $B_\phi$ show a discontinuity at the cylindrical radius 
    of about $R\sim25$ wherein there is the FMSS shock at $z = 100$. The 
    fluctuations of $B_R$ close to the axis are exaggerated because $B_R$ is of 
    small amplitude around the $z$-axis. The discontinuities of $v_z$ and $B_z$ 
    at the FMSS are too small to be visible.}
  \label{Fig_dev}
\end{figure*}

In order to test for possible deviations from axisymmetry, we use the results 
of our 2.5D runs as reference. For each grid cell in the finer 3D run, we plot 
its quantities against its cylindrical radius and compare it with profiles in 
the 2.5D runs (Fig. \ref{Fig_dev}).

Most of the profiles show a very good agreement between the 3D run and the 
axisymmetric 2.5D runs and also almost no signs of deviations from axisymmetry. 
For values of $R > 100$, we have only data points from the 3D run. In this 
region in the corners of the 3D domain, a small scatter of points around the 
mean profile is present which is possibly connected to boundary effects. 

In the range $25 < R < 80$, the profiles of the 2.5D and 3D simulations are in 
very good agreement in all quantities. For even smaller values of $R$, already 
the three 2.5D simulations with different resolutions show slightly different 
profiles, which result in different positions of the FMSS between $R=14$--24 
depending on the resolution. The position of the FMSS in the finer 3D run is 
also different to that in all three 2.5D runs. 

Inside the FMSS for radii $R > 10$, the profiles show again only small 
deviations from the axisymmetric mean profiles. In the innermost central spine 
of the flow, $R < 10$, we see two distinct effects. The first one is the 
changed behaviour in $v_\phi$ and $B_R$, which go to zero on the axis faster 
than in the 2.5D simulations (in these simulations $v_\phi$ and $B_R$ also 
vanish on the axis due to the imposed axisymmetric boundary conditions). 
The second effect is significantly increased deviations from axisymmetry which 
are mainly in the toroidal velocity $v_\phi$ and the poloidal magnetic field 
components $B_R$ and $B_z$. These three quantities show patterns of an 
instability. We note however, that these quantities are not dynamically 
important, since at $z=100$ the flow is super-fast magnetosonic and the 
dominant components of the magnetic field and flow speed are $B_{\phi}$ and 
$v_p$, respectively.

\begin{figure}
  \centering
  \includegraphics[width=0.66\columnwidth]{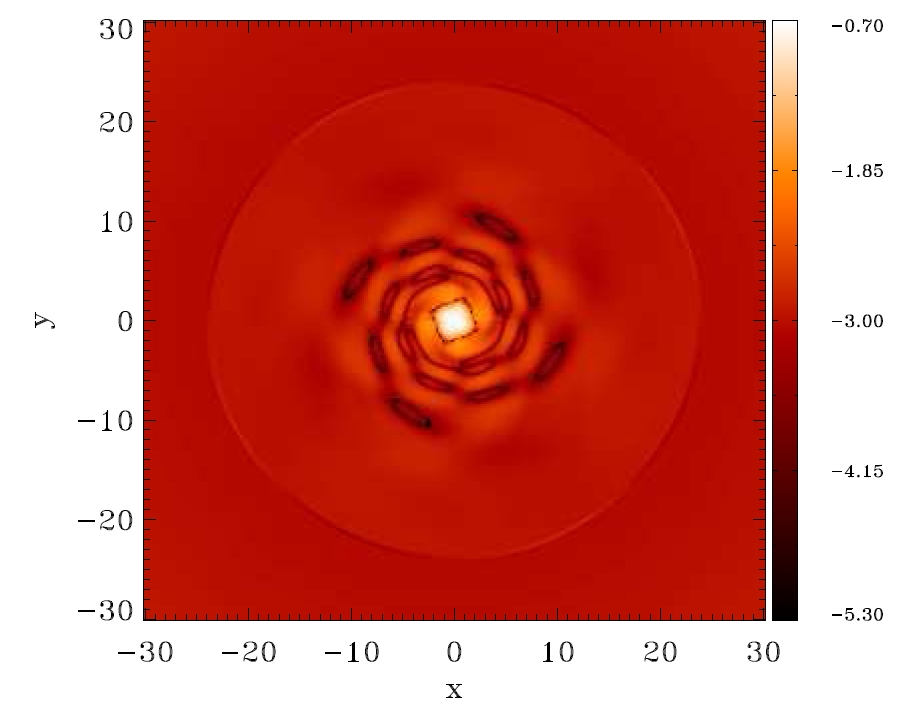}
  \includegraphics[width=0.86\columnwidth]{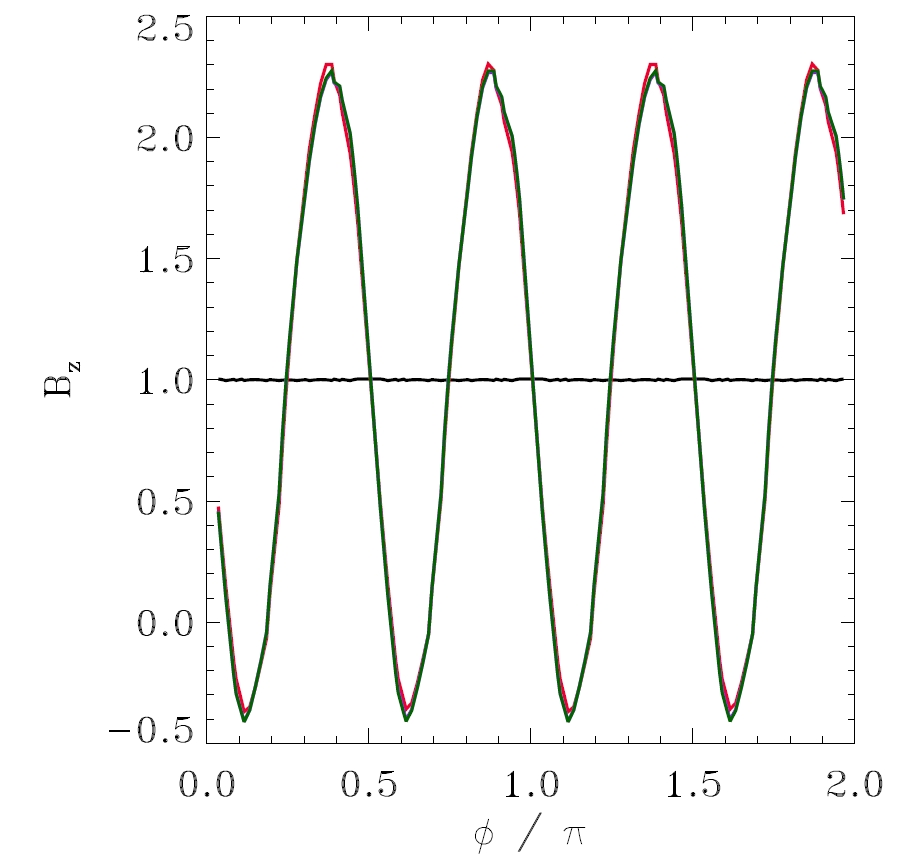}
  \includegraphics[width=0.8\columnwidth]{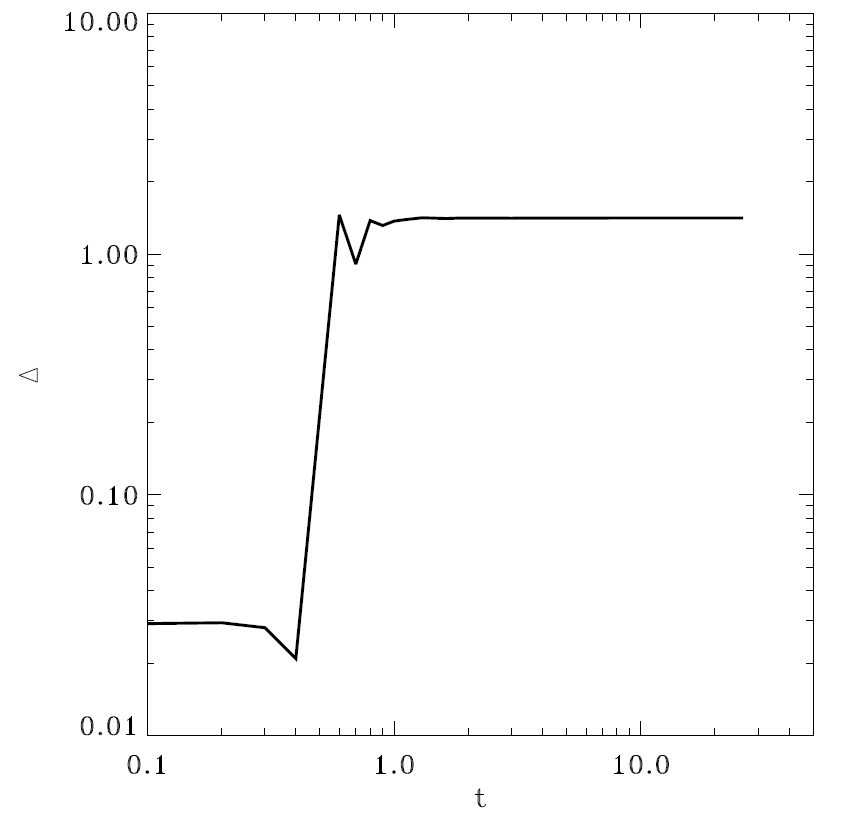}
  \caption{A horizontal slices of $\log ( | B_z | )$ at $z = 100$ in the 
    fine 3D run for time step 50 $t_0$ showing the deviations from axisymmetry 
    (top). The four peaks of the $B_z$ distribution have small amplitudes and 
    may be triggered by boundary effect due to our Cartesian domain. All 
    deviations from axisymmetry are inside the FMSS where the flow is 
    practically ballistic and therefore they do not affect the whole structure. 
    Outside the FMSS the 2.5D and 3D systems practically are identical. In the 
    middle panel, we plot $B_z$ as a function of $\phi$ normalized to the 
    average value of $B_z$ along a ring of radius $R = 7$ (and again at 
    $z = 100$) showing the sinusoidal behaviour at time steps 0 $t_0$ (black), 
    1 $t_0$ (red), 2 $t_0$ (purple) and 50 $t_0$ (green). In the bottom panel we
    show the temporal variation of the normalized amplitude, i.e. 
    $\Delta = (B_{z, \textrm{max}} - B_{z, \textrm{mean}})/B_{z, \textrm{mean}}$ as a 
    function of time.}
  \label{Fig_Bvariance}
\end{figure}

In Fig. \ref{Fig_Bvariance} (top), we show a horizontal slice of 
$\log ( | B_z | )$ through the computational domain at $z = 100$ for the final 
time step. One can directly see that the instabilities are confined to the 
interior of the FMSS, i.e.the fast-magnetosonic flow. The FMSS is fully 
developed after one orbit at 1 $t_0$ and also the instabilities do not grow 
after this time. 

When plotting $B_z$ along a ring of radius $R = 7$ (and again at $z = 100$) as 
a function of $\phi$ (Fig. \ref{Fig_Bvariance}, middle) we see a sinusoidal 
behaviour which remains steady at large times. Related to its temporal behaviour
we can distinguish between three stages (Fig. \ref{Fig_Bvariance}, bottom). Up 
to 0.5 $t_0$, the FMSS is still inside this radius and the variation of $B_z$ 
as a function of $\phi$ is only several percent. When the FMSS has expanded 
beyond the radius $R = 7$, the variance of $B_z$ rapidly grows until at 1 $t_0$ 
the instability is fully established and saturates.

Most likely, this instability is induced by the rotating flow discretized on
Cartesian grid cells. Since grid errors are replicated in the four quadrants, 
this would also explain the formation of four equidistant spiral patterns as 
seen in Fig. \ref{Fig_Bvariance}.

\subsection{Integrals of motion} \label{sec_integrals}

\begin{figure}
  \centering
  \includegraphics[width=0.85\columnwidth]{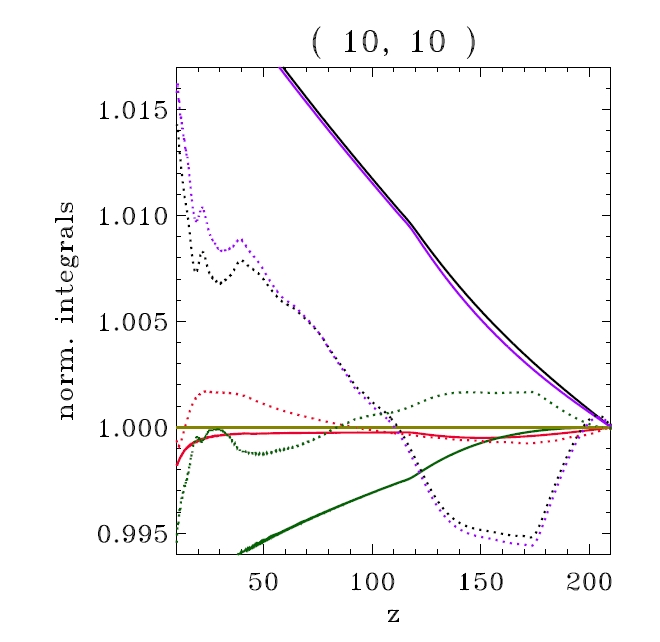}
  \includegraphics[width=0.85\columnwidth]{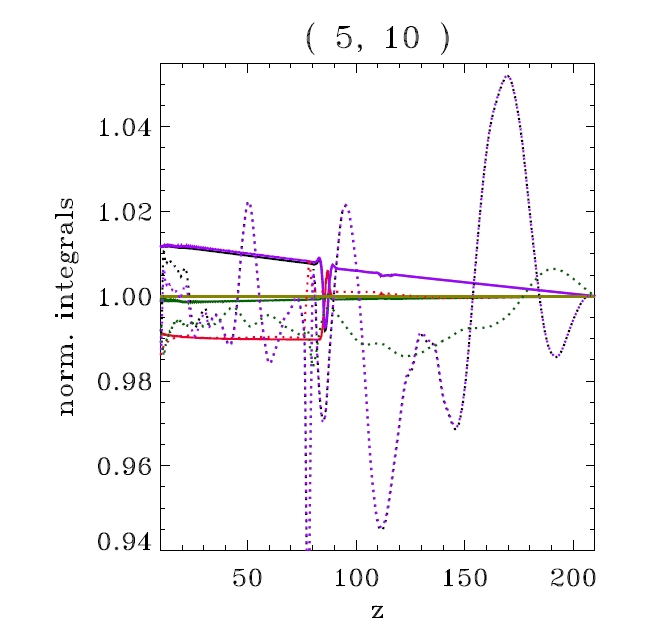}
  \includegraphics[width=0.85\columnwidth]{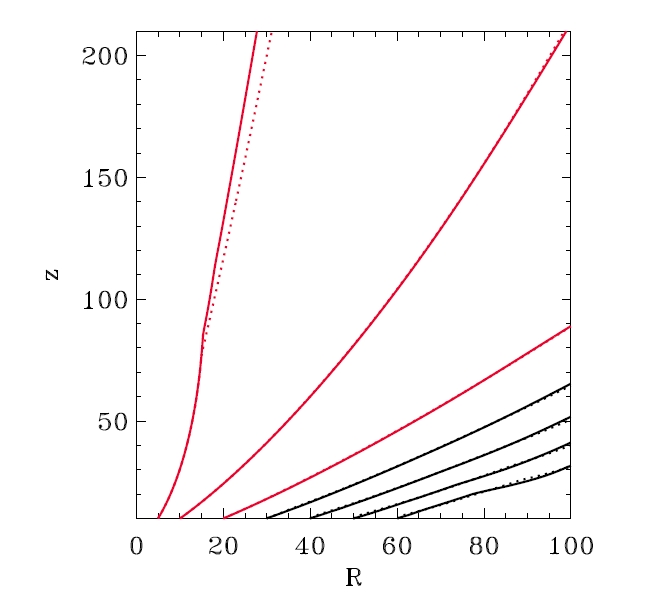}
  \caption{The integrals of motion $\Psi_A (A)$ (black), $Q (A)$ (red), 
    $\Omega (A)$ (purple), $L (A)$ (green) and $E (A)$ (yellow), respectively, 
    along surfaces with constant energy $E (A)$ anchored at the lower 
    boundary at ($R$,$z$)=(10,10) (top) and (5,10) (middle) for 
    the 2.5D run with 512$\times$1024 grid cells (solid lines) and the finer
    3D run (dashed lines). The bottom plot shows the shape of these surfaces 
    in both runs, while the three surfaces are highlighted in red, on which 
    the integrals of motion and the force components are calculated (Figs. 
    \ref{Fig_forces}--\ref{Fig_forces2}). The system reaches a steady state in 
    the 2.5D and 3D runs.}
  \label{Fig_integrals}
\end{figure}

We investigate the known five integrals of motion (see e.g., paper~I for 
details)
\begin{eqnarray} \label{integrals1}
\Psi_A (A) &=& \frac{\rho\,v_p}{B_p} \,, \\
\Omega (A) &=& \frac{1}{R}\,\left( v_\phi - \frac{\Psi_A\,B_\phi}{\rho} \right)
\,, \\
L (A) &=& R\,\left( v_\phi - \frac{B_\phi}{\Psi_A} \right) \,, \\
E (A) &=& \frac{v^2}{2} + \frac{\Gamma}{\Gamma - 1}\,\frac{p}{\rho} + \Phi - 
\Omega\,R\,\frac{B_\phi}{\Psi_A} \,, \\
Q (A) &=& \frac{p}{\rho^\gamma} 
\end{eqnarray} 
along surfaces with constant energy $E (A)$ anchored at the lower boundary at 
($R$,$z$)=(10,10) and (5,10) and $\phi = 0$, i.e. in the $y = 0$ plane in the 
finer 3D run. The values of the integrals are plotted for the 2.5D run with 
512$\times$1024 grid cells as solid lines and for the 3D run with $1024^3$ grid 
cells as dashed lines in Fig. \ref{Fig_integrals}, both for the initial state 
as well as the final state. They are normalized by their value at the upper 
boundary. Also plotted are the shapes of these surfaces which agree well in 
both runs. 

The integrals of motion directly show that we reached a steady state in both the
2.5D and 3D runs. Each integral of motion varies on a certain surface only by 
a few percent. In both runs, we find a very similar behavior of the integrals. 
Along the surface which is always inside the FMSS, $\Psi_A (A)$ (black lines) 
and $\Omega (A)$ (purple lines) show in the 3D run oscillations around a 
constant mean value. 

\subsection{Forces} \label{sec_forces}

\begin{figure*}
  \centering
  \includegraphics[width=0.85\columnwidth]{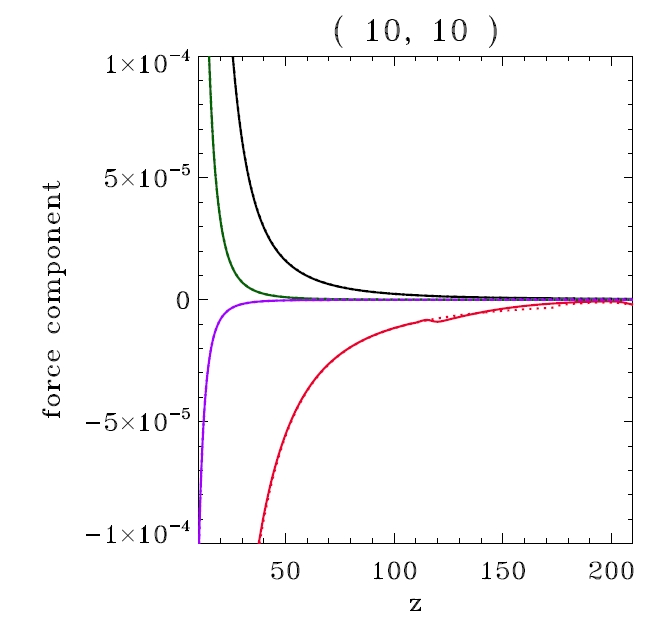}
  \includegraphics[width=0.85\columnwidth]{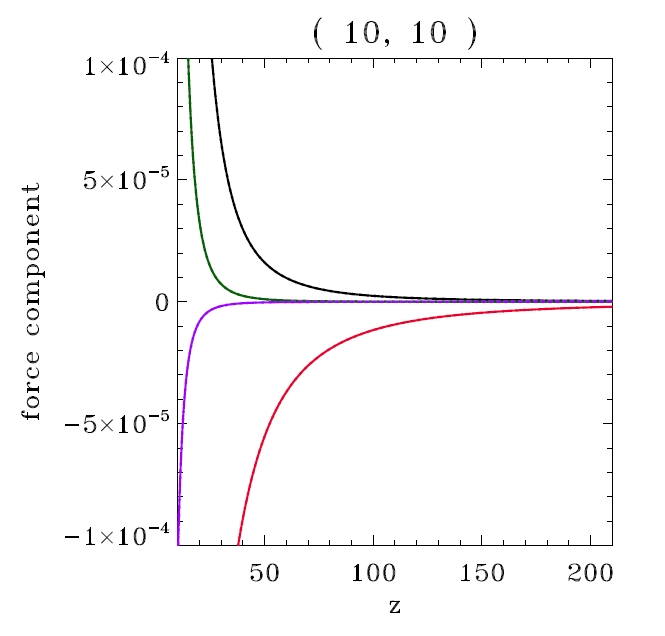} \\
  \includegraphics[width=0.85\columnwidth]{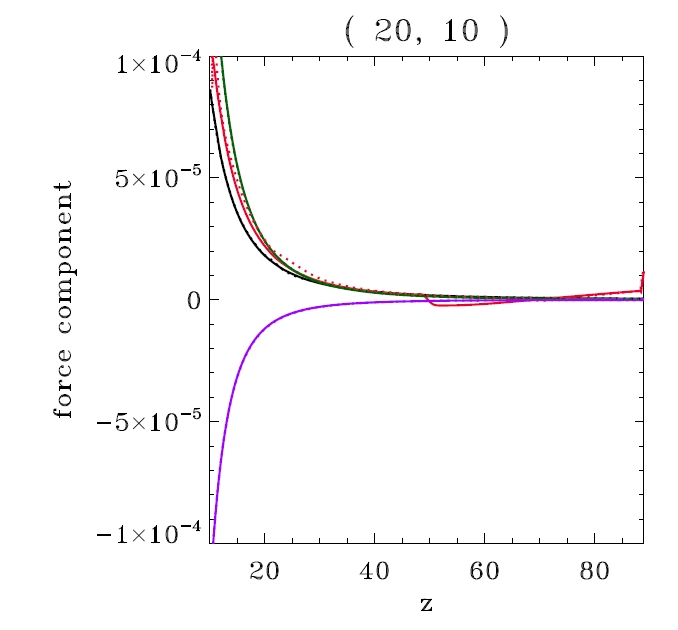}
  \includegraphics[width=0.85\columnwidth]{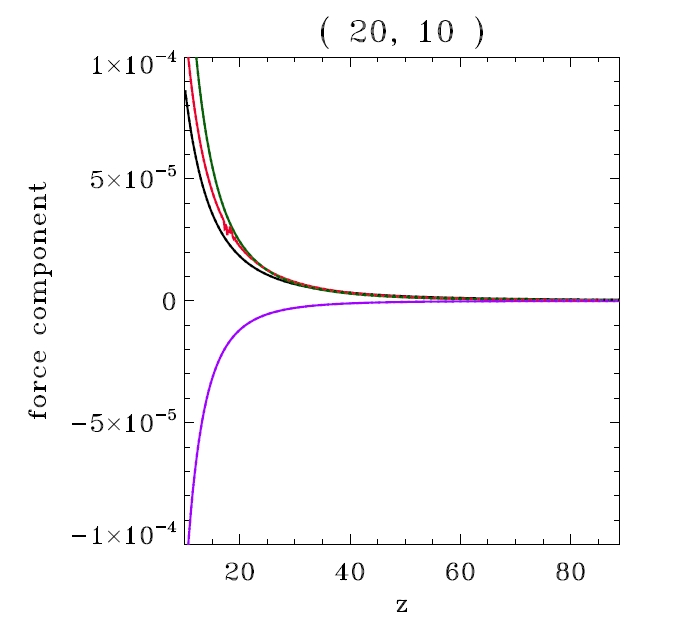}
  \caption{The $R$ components of forces along surfaces with constant energy 
    $E (A)$ anchored at the lower boundary at ($R$,$z$)=(10,10) and (20,10) 
    for the 2.5D run with 512$\times$1024 grid 
    cells (solid lines) and the 3D run (dashed lines). The colors show the 
    pressure gradient (black), Lorentz force (red), centrifugal force (green) 
    and gravity (purple). On the left is the final solution, on the right the
    initial conditions.}
  \label{Fig_forces}
\end{figure*}

\begin{figure*}
  \centering
  \includegraphics[width=0.8\columnwidth]{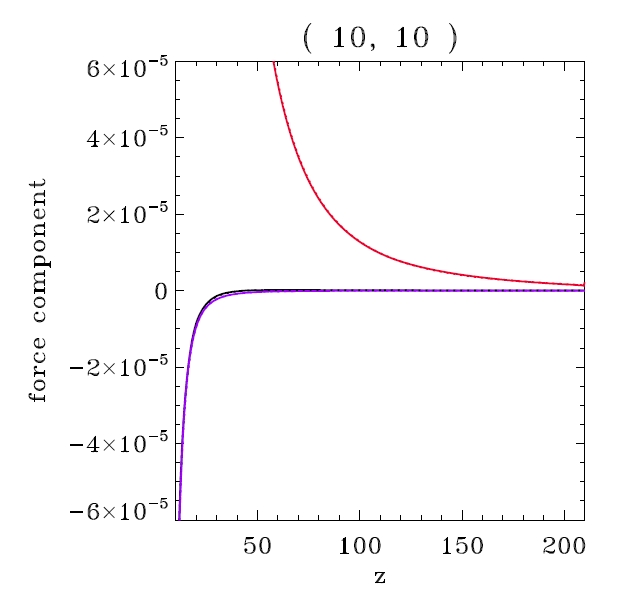}
  \includegraphics[width=0.8\columnwidth]{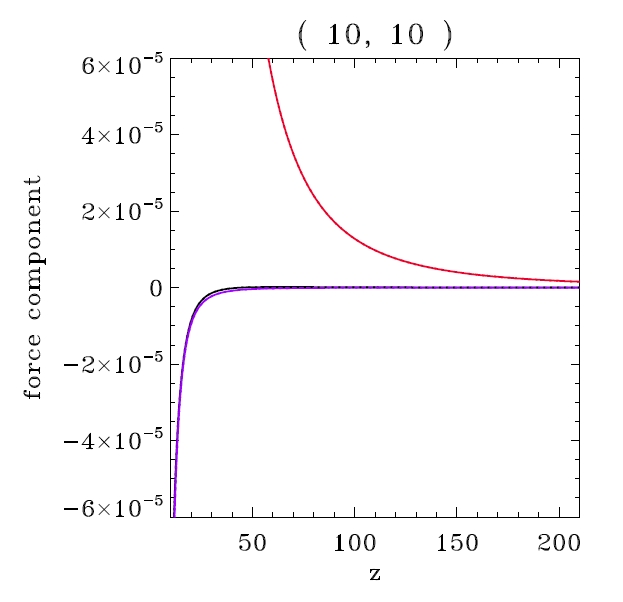} \\
  \includegraphics[width=0.8\columnwidth]{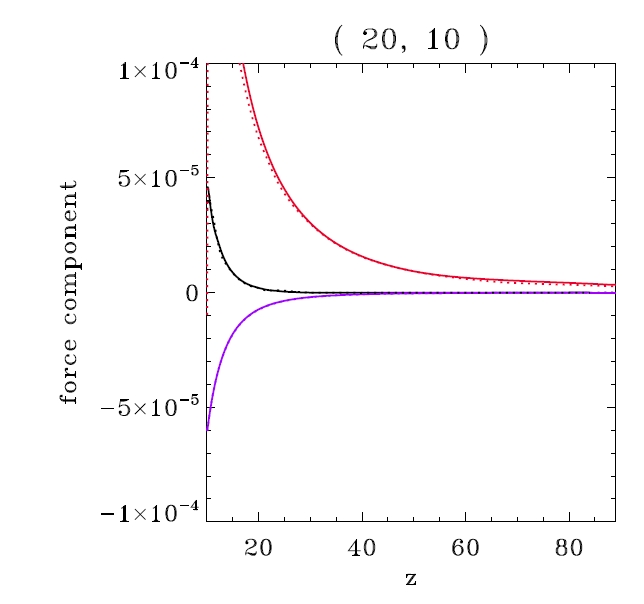}
  \includegraphics[width=0.8\columnwidth]{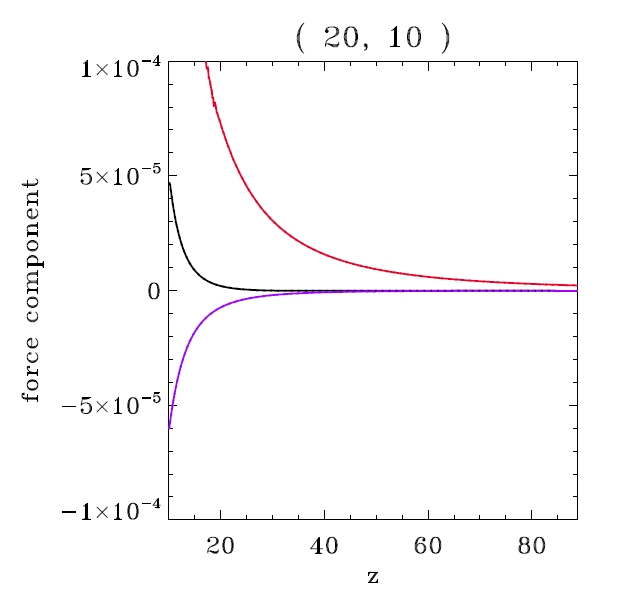}
  \caption{The $Z$ components of forces along surfaces with constant energy 
    $E (A)$ anchored at the lower boundary at ($R$,$z$)=(10,10) and (20,10) 
    for the 2.5D run with 512$\times$1024 grid 
    cells (solid lines) and the 3D run (dashed lines). The colors show the 
    pressure gradient (black), Lorentz force (red) and gravity (purple). On 
    the left is the final solution, on the right the initial conditions.}
  \label{Fig_forces2}
\end{figure*}

We also investigate the $R$ and $Z$ components of forces along the same surfaces
as in the previous section, see Figs. \ref{Fig_forces}--\ref{Fig_forces2}. 
The interplay between the $R$ components of the pressure gradient, the 
centrifugal and the Lorentz force is responsible for collimating the flow or 
triggering its expansion, their sum being related to the curvature of the 
poloidal surfaces. They also contribute to the flow acceleration, especially on 
the outer surfaces whose inclinations with respect to the vertical are not 
small. On these outer surfaces, the three forces are working against gravity. 
The Lorentz force is larger than the pressure force, however, is smaller than 
the centrifugal force up to some distance, indicating the magnetocentrifugal 
initial driving of disk-winds, and becomes larger at larger distances, showing 
the contribution of the $\bmath J \times \bmath B=(\nabla\times\bmath{B})\times 
\bmath{B}$ force to the flow acceleration. On the inner surface (which is close 
to vertical), the Lorentz force always collimates the flow. This is expected, 
since the $R$ component of the Lorentz force equals $-J_z B_\phi$ with $B_\phi<0$ 
and $J_z=\partial (RB_\phi)/(R\partial R)<0$ near the axis. On the contrary the 
``return current'' is $J_z>0$ at larger distances, resulting in negative force, 
as seen in the outer surface anchored at ($R$,$z$)=(20,10).

For the $Z$ components of the forces (Fig. \ref{Fig_forces2}) it is clear that 
the Lorentz force always dominates the pressure gradient and gravity, 
contributing to the flow acceleration.

\section{Summary and conclusions}

We have shown the results of the first 3D simulation of a disc-wind crossing 
all magnetohydrodynamic critical surfaces. This result is important in order to 
assure that the main flow is causally disconnected from its source, which is a 
prerequisite for a jet with the observed long lifetime. We have compared these 
results with previous axisymmetric 2.5D simulations.

The structure of the flow in all simulations exhibits strong similarities. In 
the outer part of the flow, its structure is almost identical in all cases. 
Near and at the position of the fast magnetosonic separatrix surface (FMSS) 
which shows as a shock, some minor deviations of the 3D run from the 2.5D runs 
are present. 

The 3D runs reach a steady state and stay close to axisymmetry for most of the 
variables, except for the poloidal magnetic field and the toroidal velocity 
which deviate considerably from axisymmetry, but are not dynamically important. 
The latter quantities show signs of instabilities, which, however, are confined 
to the region inside the FMSS. It is important to emphasize that the crossing 
of the FMSS, which represents the ``event horizon'' for the propagation of MHD 
waves, does not allow any disturbance at large distances to reach the base of 
the flow, contributing to its stability.

The forces present in the flow, both of collimating and accelerating nature, are
in good agreement in both the 2.5D and the 3D runs.

The main goal of the present paper is to check whether the axisymmetric, 
radially self-similar MHD solution for a polytropic disc-wind which crosses all 
appropriate critical surfaces to satisfy causality \citep{VTS00}, behaves 
``well'' also after (i) removing the axial singularity and (ii) relaxing the 
assumption of axisymmetry.  We have found that this solution is structurally 
stable to non-axisymmetic perturbations. The next step will be to further 
configure and improve this solution such as it may describe realistic 
astrophysical disc-winds. As in paper~I, this further extension will involve 
the truncation of the solution at some arbitrary radii and use of this 
truncated solution to study the temporal evolution of jets within an MHD model 
and further comparison with observations. 
\citet{SNO10} argue that models exhibiting a slowly varying poloidal field 
component in the accretion disc (their model OP) match observations better than 
those resembling self-similar MHD solutions (as their model BP). Note, however, 
that in paper~II we showed that radially truncated self-similar solutions may 
very well describe real HST observations.

\section*{Acknowledgments}
We acknowledge the referee, Marina Romanova, for helpful comments and 
suggestions that improved the paper. The 3D simulations have been performed on 
the supercomputers BCX at CASPUR, on SP6 at CINECA and on supercomputers in the 
bwGRiD, the grid of the state Baden W\"urttemberg.

\bsp

\label{lastpage}
\end{document}